\documentstyle[10pt]{article}
\begin{document}
\setlength{\baselineskip}{0.9 cm}

\vskip 0.3in
\centerline{\bf \large The Hydrogen Atom as an Entangled Electron-Proton System}
\vskip 0.25in
\centerline{Paolo Tommasini and Eddy Timmermans}
\centerline{Institute for Theoretical Atomic and Molecular Physics}
\centerline{Harvard-Smithsonian Center for Astrophysics}
\centerline{Cambridge, MA 02138, USA}
\vskip 0.15in
\centerline{A.F.R. de Toledo Piza}
\centerline{ Instituto de Fisica, Universidade de S\~ao Paulo}
\centerline{C. P. 66318, S\~ao Paulo, SP 05315-970, Brazil}

\vspace{2cm}

\centerline{                ABSTRACT}
We illustrate the description
of correlated subsystems
by studying the simple two-body Hydrogen atom.
We study the entanglement of the electron and proton coordinates
in the exact analytical solution.  This entanglement, which
we quantify in the framework of the density matrix formalism,
describes correlations
in the electron-proton motion.
\newpage
\section{Introduction}

        Independent particle models often provide a good starting
point to describe the physics of many-particle systems.  In these
models, the individual particles behave as independent particles that
move in a potential field which accounts for the interaction with the
other particles in an average sense.  However, depending on the systems and the
properties that are studied, the results of the independent particle
calculations are not always sufficiently accurate.  In that case it is
necessary to correct for the fact that the motion of a single particle
depends on the positions of the other particles, rather than on some
average density.  Consequently, in a system of interacting particles,
the probability of finding two particles with given positions or
momenta is not simply the product of the single particle
probabilities: we say that the particles are `correlated'.

        In a more general context, the problem can be formulated as
the description of interacting subsystems (the single particle in the
many-particle system being the example of the subsystem).  A
convenient theoretical framework to deal with this problem in quantum
mechanics is provided by the density matrix theory, which is almost as
old as quantum mechanics itself \cite{vone, esch}.  The ever-present
interest in density-matrix theory, in spite of its long history, is
justified by the power of its description and the importance of its
applications, which extends to the study of the very foundations
of Quantum Theory \cite{vone, esch, omn}.  In this article, we
briefly review some important aspects of density matrix theory and we
illustrate its use for describing correlations of interacting
subsystems by studying the simple, exactly solvable system of the
Hydrogen atom.

\section{Subsystems and Quantum Correlations}

In very broad terms, giving the state of a physical system means
providing the necessary information to evaluate all its observables
quantitatively. The states of a quantum mechanical system are
frequently represented by vectors of unit norm in a Hilbert
space ${\cal H}$, in general of infinite dimension (hence their
usual designation as 'state vectors'). If the system is in a state
$|\Psi \rangle$, with $\langle\Psi |\Psi\rangle=1$, the outcome of
measurements of an observable quantity A has expectation value
$\langle A \rangle$ equal to
\begin{equation}
\langle A \rangle = \langle \Psi | \hat{A} | \Psi \rangle \; \; \; ,
\label{e:exp1}
\end{equation}
where we use a hat, `$\hat{A}$', to distinguish the operator
representing the observable from the observable itself.  If the
vectors $|w_{i} \rangle$, $i=1,2,...$  are a basis of the Hilbert space
${\cal H}$, we can use the completeness relation, $1 = \sum_{m}
|w_{m}\rangle \langle w_{m}|$, to write the expectation value as
\begin{equation}
\langle A \rangle = \sum_{m,n} \langle \Psi | w_{m} \rangle
\langle w_{m} | A | w_{n} \rangle \langle w_{n} | \Psi \rangle \; \; .
\label{e:exp2}
\end{equation}
In matrix notation, the $m n$-matrix element of the $\hat{A}$-operator
in the $|w\rangle$-basis is equal to $\hat{A}_{m n}= \langle w_{m}
|\hat{A}| w_{n} \rangle$. The remaining ingredients of
Eq. (\ref{e:exp2}) can be collected as $\rho_{n m} = \langle w_{n} |
\Psi \rangle \langle \Psi | w_{m} \rangle$, known as the density
matrix \cite{vone, landau, fano}, which is recognized as the matrix
representing the so called density operator, $\rho=|\Psi\rangle
\langle \Psi|$, in the chosen basis. The expectation value of
Eq. (\ref{e:exp2}) can then be written as
\begin{equation}
\langle A \rangle = \sum_{m,n} \hat{A}_{m n} \rho_{n m} \; \; ,
\label{e:exp3}
\end{equation}
which is the trace of the product of the $A$ and $\rho$-matrices,
$\langle A \rangle = Tr(\hat{A}\rho)$.  In this expression, the
state of the system is represented by the density matrix or,
equivalently, by the density operator $\rho$. The form of this
operator, together with normalization property of the underlying state
vector $|\Psi\rangle$, lead to some important properties, which are of
course shared by the corresponding density matrices. Namely, the
density operator is a non-negative (meaning that
$\langle\phi|\rho|\phi\rangle\geq 0$ for any vector $|\phi\rangle$)
self-adjoint operator ($\rho=\rho^\dagger$), of unit trace
($Tr(\rho)=1$), and idempotent (meaning that $\rho=\rho^2$, which 
follows from $|\Psi \rangle \langle \Psi | \Psi \rangle \langle \Psi | =
| \Psi \rangle \langle \Psi| $).

The use of state vectors $|\Psi\rangle$ to describe the state of a
quantum system is however not general enough to cover many frequently
occuring situations. As we shall discuss in continuation, a suitable
generalization is provided by the density matrix language, if only we
relax the idempotency requirement.

        Often, when probing a system, only part of the total system is
subjected to a measurement.  For example, in a typical scattering
experiment only the scattered particle is detected. The target system
is left behind in a final state that could be different from its
initial state, but this state is not observed directly. It is then
natural to divide the total system into the target system and the
single particle system of the scatterer.  Also, in many
experiments involving mesoscopic, or semi-macroscopic, quantum
devices, it is very hard, if not impossible, to ensure efficient
isolation from the environment, which then plays the role of a second
subsystem coupled to the system of interest \cite{kabro}.
Mathematically, such partitioning of the system into subsystems means
factorizing the Hilbert space ${\cal H}$ into the tensor product of
the corresponding subspaces, ${\cal H} = {\cal H}_{u} \otimes {\cal
H}_{v}$.  If the ${\cal H}_{u}$ and ${\cal H}_{v}$ -spaces are spanned
respectively by the bases $|u_{i}\rangle$, $i=1,2,...$ and
$|v_{j}\rangle$, $j=1,2,...$, then ${\cal H}$ is spanned by the
product states $|u_{i}\rangle |v_{j}\rangle$.  Accordingly, a
state vector $|\Psi \rangle$ describing the state of the system can
be expanded as
\begin{equation}
|\Psi \rangle = \sum_{i,j}d_{i,j} |u_{i}\rangle |v_{j} \rangle \; \; .
\end{equation}

Suppose now that one wishes to describe the state of one of the
subsystems alone, say subsystem $u$. The observable quantities
of this subsystem are represented by operators which act non-trivially
on the vectors of ${\cal H}_u$ while acting on the vectors of ${\cal
H}_v$ simply as the unit operator.  The expectation value of such
operators can be obtained from the
reduced density matrix  for the $u$-system,
which is defined as the trace of the full density
matrix $\rho$ over the $v$-subspace. When the entire system is in
the state described by $|\Psi\rangle$ this reads
\begin{eqnarray}
\rho_{u} &=& Tr_{v} \left( \sum_{i,j} \sum_{k,l}
| u_{i}\rangle | v_{j} \rangle d_{i,j}
d^{\ast}_{k,l} \langle u_{k}| \langle v_{l} | \right)
\nonumber \\
&=& \sum_{i,k} \sum_{j} |u_{i} \rangle
d_{i,j} d^{\ast}_{k,j} \langle
u_{k} | 
 \; \; .
\label{e:rhou}
\end{eqnarray}
We can understand the trace 
as a sum $\sum_{j}$ over the probabilities for the $v$-subsystem to 
be in any possible $|v_{j}\rangle$-state.
For example, in the scattering experiment where only
the state of the detected particle is determined, the
final state of the target ($v$)-system is not measured, and one 
has to sum over the probabilities of finding the target in
all possible $|v_{l}\rangle$-states. 

An important limiting situation is that in which the overall state
vector $|\Psi\rangle$ is itself a product of a $u$-vector and a
$v$-vector, $|\Psi \rangle = |u\rangle \; |v\rangle$.  
For these product states, 
the density matrix $\rho = |\Psi\rangle \langle \Psi|$,
factorizes,
\begin{equation}
\rho = |u\rangle |v\rangle \; \langle v| \langle u|=
\rho^{u} \; \rho^{v} \; \; ,
\label{e:nonin}
\end{equation}
and the reduced density $\rho^u$ is simply given by
$|u\rangle\langle u|$.  This is just of the form of a density operator
associated with the state vector $|u\rangle$, which can then also be
used to describe the state of the subsystem. In general, the state
vector $|\Psi\rangle$ cannot be factorized in this fashion,
however. In this case one says that the two subsystems are in an
`entangled' state.  The density matrix of an entangled state does not
factorize and appears as a series of $u$ and $v$-products: the
subsystems are correlated.  The number of terms gives some indication
of the departure from the uncorrelated density matrix, but as the
number of terms depends on the choice of basis in the $u$ and
$v$-spaces, this is not a good measure of correlation or
entanglement. A way out of this difficulty is however provided by
a simple and remarkable result due to Schmidt \cite{schm} (see also
\cite{esch, ric, kni2}) which essentially identifies a `natural' basis (in
the sense of being determined by the structure of the entangled state
itself) in which the description of the entanglement is achieved with
maximum simplicity.

The Schmidt basis is a product basis, in the usual sense that it is
written as the tensor product of two particular bases, one for the
${\cal H}_u$ space and another for the ${\cal H}_v$ space. In order to
find these two bases one looks for the eigenvectors of the reduced
density matrices.  If the state is not entangled, $|\Psi\rangle =
|u\rangle |v\rangle$, $\rho^{u}$ and $\rho^{v}$ each have one single
eigenvector of non-zero eigenvalue, $|u\rangle$ for $\rho^{u}$ and
$|v\rangle$ for $\rho^{v}$, and the numerical value of the
corresponding eigenvalues is equal to 1.  One then
completes the basis in each subspace by including enough additional
orthonormal vectors which are moreover orthogonal to the single
'relevant' vector with nonzero eigenvalue. Each one of these aditional
basis vectors is then an eigenvector of the corresponding reduced
density with eigenvalue zero, and the large degree of arbitrariness in
choosing them reflects the large degeneracy of the
corresponding eigenvalue zero.  If, on the other hand, the subsystems
are correlated, then $\rho^{u}$ and $\rho^{v}$ have a spectrum of 
nonvanishing eigenvalues.  Interestingly, as we prove below, the
eigenvalues of the $\rho^{u}$ and $\rho^{v}$-operators are the
same. To see that, we assume that the $\rho ^{u}$-matrix in the
$|u_{i}\rangle$ basis, is diagonalized by the unitary transformation
$U$ ($ \sum_{i} U_{i l} U^{\ast}_{i j} = \delta_{l j} $; $\sum_{i}
U_{l i} U^{\ast}_{j i} = \delta_{j l}$),
\begin{equation}
\sum_{k} \rho^{u}_{i k} U_{k l} = \lambda_{l} U_{i l} \; \; .
\label{e:rhoud}
\end{equation}
The $\rho^{u}$-matrix elements are given by Eq.(\ref{e:rhou}),
$\rho^{u}_{i k} = \sum_{j} d_{i j} d^{\ast}_{k j}$, so that 
Eq.(\ref{e:rhoud}) can be written as
\begin{equation}
\sum_{k,j} d_{i j} d^{\ast}_{k j} U_{k l} = \lambda_{l} U_{i l} \; \; .
\label{e:rhoud2}
\end{equation}
Multiplying Eq.(\ref{e:rhoud2}) by $d^{\ast}_{i l'}$, and summing 
over $i$, we obtain
\begin{equation}
\label{c}
\sum_{i,j} \sum_{k} d^{\ast}_{i l'} d_{i j}
d^{\ast}_{k j} U_{k l} = \lambda_{l}
\sum_{i} d^{\ast}_{i l'} U_{i l}
\label{e:rho3}
\end{equation}
Defining $V_{j,l} = \sum_{k} d^{\ast}_{k j} U_{k l}$, 
we find with Eq.(\ref{e:rho3}) that the vector of components
$(V_{1l},V_{2l},\ldots,V_{jl},\ldots)$ is an eigenvector of
$\rho^{v}$ with eigenvalue $\lambda_{l}$:
\begin{equation}
\sum_{j} \rho^{v}_{l' j} V_{j l}= \lambda_{l} V_{l l'} \; ,
\label{e:rho4}
\end{equation}
where $\rho^{v}_{l' j} = \sum_{i} d^{\ast}_{i l'} d_{i j}$.  
It follows that
$\rho^{u}$ and $\rho^{v}$ have
the same eigenvalues $\lambda$. If the nonvanishing eigenvalues
are not degenerate, then all these vectors are automatically
orthogonal. And if, moreover, the set of eigenvectors with
nonvanishing eigenvalue is still not complete, one still may complete
the bases with additional orthogonal vectors which are again
eigenvectors of the respective reduced densities with eigenvalue
zero. 
Expanding $|\Psi \rangle$ in the product basis constructed in this
way, which thus includes the eigenvectors $|u_{\lambda}
\rangle$ of $\rho^{u}$ and $|v_{\lambda}\rangle$ of
$\rho^{v}$,$
|\Psi \rangle = \sum_{\lambda} c_{\lambda} |u_{\lambda} 
\rangle |v_{\lambda} \rangle, $ 
the reduced density matrices take on the form
\begin{eqnarray}
\rho^{u} &=& \sum_{\lambda}  |c_{\lambda}|^{2} |u_{\lambda}\rangle 
\langle u_{\lambda}|   \;  ,
\label{e:rho5}
\end{eqnarray}
which is diagonal and shows that the $\lambda$-eigenvalues are equal
to $\lambda = |c_{\lambda}|^{2}$.  This also shows that the
eigenvectors with zero eigenvalue do not participate in the
expansion.  The eigenvalues $\lambda$ are thus both the
probabilities of finding subsystem $u$ in the states
$|u_\lambda\rangle$ and the probabilities of finding subsystem $v$ in
the states $|v_\lambda\rangle$. The set of numbers $|c_{\lambda}|^{2}$
can therefore be interpreted as a distribution of occupation
probabilities. In fact, the unit trace condition on the complete
density operator $\rho$ ensures that $\sum_\lambda |c_\lambda|^2=1$.
For uncorrelated subsystems, there is just one nonvanishing
probability and this distribution has a vanishing `width' or standard
deviation.  It is then reasonable to use the standard deviation as a
measure of the correlation \cite{entro}. Note that when the two
subsystems are correlated, i.e., there is more than just one
nonvanishing occupation probability $|c_{\lambda}|^{2}$, the reduced
density matrix Eq. (\ref{e:rho5}) is no longer idempotent, although
all the other properties listed above for $\rho$ still hold. As a
result of this, the state of an entangled subsystem cannot be
described in terms of a state vector in the corresponding Hilbert
space, and one is forced to use the density matrix language. In this
case one says that the subsystem is in a `mixed' state. `Pure' states,
associated with definite state vectors, are on the other hand always
associated with idempotent density operators.

In summary, the foregoing discussion shows that a
subsystem of a larger quantum system is, in general, in a mixed state,
even if the state of the system as a whole is pure. Furthermore, the
Schmidt analysis shows that in this case the state of the
complementary subsystem has exactly the same degree of impurity,
in the sense that both reduced density matrices have the same
spectrum. This corresponds to the impurity of both subsystems being
due to their mutual entanglement in the given overall pure state.  To
the extent that a given quantum system is not really isolated, but
interacts with other systems (possibly a non-descript `environment'),
this analysis also shows that assuming the purity of its state is
inadequately restrictive. In the following section we
examine the spatially homogeneous, but possibly not pure, states of a
single quantum particle, independently of any dynamical processes that
may in fact give real existence to such states. An example of such a
dynamical process will be given next. 

\section{Single Particle System}

        For the sake of illustration, we apply the density matrix
description to a single, 'elementary' particle system. In spite
of the extreme simplicity of this type of system, the density matrix
language enables us to consider states that are not accessible within
the standard state vector (or wavefunction) description - mixed states
- which are of interest for the discussion of correlated many-particle
systems.

In the coordinate representation, which corresponds to the choice of
the eigenfunctions $|{\bf r}\rangle$ of the coordinate operator as
basis, the density matrix $\rho$ of a pure single particle system 
represented by the state vector 
$|\Psi \rangle$, takes on the form
\begin{equation}
\rho({\bf r},{\bf r}') = \langle {\bf r} | \Psi \rangle
\langle \Psi | {\bf r}'\rangle = \Psi ({\bf r}) \Psi^{\ast} ({\bf r}')
\; \; \; ,
\label{e:denscr}
\end{equation}
where $\langle {\bf r}| \Psi \rangle = \Psi ({\bf r})$ is the wavefunction.
In the general case, 
the single particle density matrix is an object
$\rho({\bf r},{\bf r}')$
with the properties of hermiticity, $\rho({\bf r},{\bf
r}')=\rho^*({\bf r}',{\bf r})$, unit trace, $\int d^3r\rho({\bf
r},{\bf r})=1$, and non-negativity, $\int d^3r \int d^3r'\phi^* ({\bf
r})\rho({\bf r},{\bf r}')\phi({\bf r}') \geq 0$.  Note that idempotency
is not being required in order to allow for mixed states.

        We shall be concerned with a particular class of states
which we shall refer to as homogeneous states.  By definition, these
states are translationally invariant, meaning that translation
by an arbitrary position vector ${\bf R}$ leaves the
density matrix, and hence the state itself, unaltered: $\rho({\bf
r}+{\bf R},{\bf r}'+{\bf R})=\rho({\bf r},{\bf r}')$. As a
consequence, all the density matrix elements depend only on relative
positions, specifically, $\rho({\bf r},{\bf r}') = \rho({\bf r}-{\bf
r}')$. In this case, the Fourier transform $\tilde{\rho}({\bf k})$ of
$\rho({\bf r}-{\bf r}')$,
\begin{equation}
\rho ({\bf r}-{\bf r}') = \frac{1}{(2 \pi)^{3}} \int d^{3} k \;
\tilde{\rho}({\bf k}) \exp(i{\bf k}\cdot [{\bf r}-{\bf r}']) \; ,
\label{e:ft}
\end{equation}
plays a dual role in the description of the homogeneous state:
(1) it is proportional to the momentum distribution of the system, and
(2) it gives the eigenvalues of the density matrix.  We shall
now prove both assertions.

Let us first evaluate the expectation value $\langle {\bf p}
\rangle$ of the momentum.  If the system is pure and described by
the quantum state $|\Psi \rangle$, then  
\begin{eqnarray}
\langle {\bf p} \rangle &=& \int d^{3} r \; \Psi^{\ast} ({\bf r}) \;
[-i\hbar \nabla ] \; \Psi ({\bf r}) 
\nonumber \\
&=& -i\hbar \int d^{3} r \; d^{3} r' \; \delta ({\bf r} - {\bf r}') 
\nabla_{\bf r} \rho({\bf r},{\bf r}') \; \; .
\label{e:mom}
\end{eqnarray}
The trace prescription for calculating $\langle {\bf p} \rangle$,
obtained above, is also valid for mixed states and inserting Eq.(\ref{e:ft}),
we obtain
\begin{eqnarray}
\langle{\bf p}\rangle &=& -i\hbar\int d^3r\int d^3r'\delta({\bf
r}-{\bf r}')\nabla_{\bf r}\rho({\bf r},{\bf r}') \nonumber \\
&=&\frac{1}{(2\pi)^3}\int d^3r\int d^3r'\int d^3k\; [ \hbar{\bf k}] \;
\tilde{\rho}({\bf k}) \exp(i{\bf k}\cdot [{\bf r}-{\bf r}'])\;
\delta({\bf r}-{\bf r}') 
\\ &=& \frac{\Omega}{(2\pi)^3}\int
d^3k\; [\hbar{\bf k} ] \; \tilde{\rho}({\bf k}) \; ,
\nonumber
\label{e:mom2}
\end{eqnarray}
where we have used a large quantization volume
$\Omega$ to avoid problems related to states of infinite norm.
This normalization scheme, commonly called box normalization, assumes that the
particle is confined to a large
volume $\Omega$, which is taken to be infinite at the end of the calculation,
$\Omega \rightarrow \infty$.   A plane
wave state is then normalized as $\exp (i {\bf k} \cdot {\bf r}) / \sqrt{\Omega}$.
Substituting ${\bf p} = \hbar {\bf k}$ in the integration of
Eq.(\ref{e:mom2}), we find that
\begin{equation}
\langle {\bf p} \rangle =
\int d^{3} p \; {\bf p} \; f({\bf p}) 
\label{e:dist}
\end{equation}
where $ f({\bf p}) = [\Omega/(2\pi\hbar)^{3}] \tilde{\rho}({\bf p}/\hbar)$.
Similarly, the expectation value of higher-order powers of 
momentum components are equal to the corresponding higher-order
moments of the $f({\bf p})$-function,
e.g. $\langle (\hat{\bf p}\cdot \hat{\bf x})^{2} \rangle = 
\int d^{3} p \; ({\bf p}\cdot \hat{\bf x})^{2} \; f({\bf p})$.  
Thus, $f({\bf p})$ may be interpreted
as a momentum distribution function. 

        To see that $\tilde{\rho}({\bf k})$ gives the
eigenvalues of the density matrix, we use Eq. (\ref{e:ft}) to
evaluate
\begin{eqnarray}
\label{e:eig}
\int d^{3} r' \; \rho({\bf r}-{\bf r}') \frac{\exp(i{\bf k}\cdot {\bf
r}')} {\sqrt{\Omega}}&=&\frac{1}{(2\pi)^3}\int d^3k'\tilde{\rho}({\bf
k}')\frac{\exp(i{\bf k}'\cdot{\bf r})}{\sqrt{\Omega}}\int d^3r'
\exp(i[{\bf k}-{\bf k}']\cdot {\bf r}') \nonumber \\
&=&\tilde{\rho}({\bf k}) 
\frac{\exp(i{\bf k}\cdot{\bf r})}{\sqrt{\Omega}}
\end{eqnarray}
which we recognize as the eigenvalue equation of the density matrix
(Eq.(\ref{e:rhoud})) in coordinate representation.  From
Eq.(\ref{e:eig}), it follows that the eigenvectors of the density
matrix are the plane wave states of momentum ${\bf k}$ and that the
corresponding eigenvalues are equal to $\tilde{\rho}({\bf k})$.

	The density matrix of a homogeneous single particle
system does generally {\it not} describe a pure system.  To see that,
we remind the reader that a pure system is characterized by
an idempotent density matrix, $\rho^{2} = \rho$.
For the homogeneous
single particle system, this implies that the eigenvalues of the density matrix 
satisfy $\tilde{\rho}({\bf k})^{2} =
\tilde{\rho}({\bf k})$.  Consequently, for a pure homogeneous
single particle system, we find that $\tilde{\rho}({\bf k}) = 0
$ or $\tilde{\rho}({\bf k}) = 1$.  Furthermore,
the unit
trace condition on the density matrix implies
\begin{equation}
1=\int d^3r\;\rho(0)=\rho(0)\Omega=\frac{\Omega}{(2\pi)^3}\int d^3k
\;\tilde{\rho}({\bf k})\rightarrow\sum_{\bf k}\tilde{\rho}({\bf k}) \; ,
\end{equation}
where in the last step we reconverted the momentum integral to the sum
of discrete momenta appropriate to the adopted volume
quantization.  Thus,  the homogeneous single particle system that is pure
can only be described by a density matrix with
a single eigenvalue, say $\tilde{\rho}({\bf k_1})$, equal to $1$, and
all other eigenvalues $\tilde{\rho}({\bf k}) = 0, {\bf k} \neq {\bf k}_{1}$.
Such density matrix represents a particle
with a definite momentum (zero standard deviation of the distribution
of density matrix eigenvalues), corresponding to a plane wave 
wavefunction, $\langle{\bf r}|\Psi\rangle = \exp(i{\bf k}_{1}
\cdot {\bf r})/\sqrt{\Omega}$.  In contrast, an impure single
particle system, has a spectrum $\tilde{\rho}({\bf k})$ of density
matrix eigenvalues.  The standard deviation of the $\tilde{\rho}
({\bf k})$-distribution is a measure of the correlation or 
entanglement of the single-particle system \cite{fin}.

It should be remarked, finally, that all the above discussion had a
purely kinematical character, in the sense that we dealt with possible
states of a quantum mechanical system and with their inherent
obervable properties, without regard to any dynamical processes that
might have led to their preparation. In the following section we
consider a simple but definite composite dynamical system and use the
tools developed here in order to study the entanglement effects
generated by definite interaction processes.

\section{Hydrogen Atom}

	We now apply the density matrix formalism to stationary
states of a system that
can display correlation: the
two-body hydrogen atom.  The hydrogen atom consists
of an electron and a proton, interacting by means of the
Coulomb potential.  The hamiltonian of the electron-proton system is
\begin{equation}
\label{x}
\hat{H} = \frac{\hat{{\bf p}}^{2}_{\rm e}}
{2 m_{\rm e}} + \frac{\hat{{\bf
p}}^{2}_{\rm p}}{2 m_{\rm p}} - \frac{e^{2}}
{|{\bf r}_{\rm e} - {\bf r}_{\rm
p}|} \; \; ,
\end{equation}
where $\hat{{\bf p}}$ and ${\bf r}$ denote the
momentum and position operators in coordinate 
space, and the $_{e}$ and $_{p}$-subscripts
indicate the electron and the proton.

	In fact, the hydrogen atom can be treated as two uncorrelated
subsystems. Transforming to relative, ${\bf r}={\bf r}_{p}
-{\bf r}_{e}$, and center-of-mass coordinates, ${\bf R} = 
(m_{e} {\bf r}_{e} + m_{p} {\bf r}_{p})/M$, where $M$ is the
total mass, $M = m_{e} + m_{p}$, the
hamiltonian separates into a center-of-mass term which
takes on the form of a free-particle hamiltonian,
$\hat{H}_{\rm cm} = \hat{p}_{R}^{2}/2M$, and an `internal' hamiltonian
which governs the relative motion of the electron and the proton,
$\hat{H}_{\rm int} = \hat{p}_{r}^{2}/2m_{r} - e^{2}/r$, where
$m_{r}$ is the reduced mass, $m_{r}^{-1} = m_{e}^{-1} + m_{p}^{-1}$.
The product of $\hat{H}_{\rm cm}$ and $\hat{H}_{\rm int}$-eigenstates,
\begin{equation}
\Psi ({\bf R},{\bf r}) = \phi_{\rm int} ({\bf r}) 
\exp(i\frac{\bf P}{\hbar} \cdot {\bf R}) /\sqrt{\Omega}
\label{e:psrr}
\end{equation}
is an eigenstate. According to the definitions introduced in section 2, 
the internal and center-of-mass subsystems for
the state (\ref{e:psrr}) are
not correlated.  If the hydrogen atom is in its atomic
ground state, the internal wavefunction, $\phi_{\rm int}$, is equal to
\begin{equation}
\phi_{\rm int}({\bf r}) = \frac{1}{\sqrt{\pi}} \left( \frac{1}{a_{o}} \right)^{3/2}
\exp(-r/a_{0}) \; \; \; \; \; \; \; \; 
({\rm hydrogen \; in \; 1s-state}) \; \; ,
\label{e:1s}
\end{equation}
where $a_{0}$ is the Bohr-radius, $a_{0} = \frac{\hbar^{2}}{m_{r} e^{2}}$.

	Partitioning the hydrogen atom differently into electron
and proton subsystems
gives an entangled state:
\begin{equation}
\Psi({\bf r}_{e},{\bf r}_{p})
= \phi_{\rm int}({\bf r}_{e} - {\bf r}_{p}) \; 
\exp \left( i \frac{{\bf P}}{\hbar} \cdot \frac{ [ m_{e} {\bf r}_{e}
+ m_{p} {\bf r}_{p} ] }{M} \right)  / \sqrt{\Omega}\; \; .
\label{e:psrerp}
\end{equation}
The reduced density matrix for the electron system can be
obtained by taking the trace over the proton basis set of coordinate
eigenfunctions:
\begin{eqnarray}
\rho({\bf r}_{e},{\bf r}'_{e}) &=&
\int d^{3} r_{p} \; \Psi({\bf r}_{e},{\bf r}_{p}) \Psi^{\ast}
({\bf r}'_{e},{\bf r}_{p}) \; \; \;
\nonumber \\
&=& \exp \left( i \frac{m_{e}{\bf P}}{\hbar M} \cdot
[{\bf r}_{e} - {\bf r}'_{e} ] \right) 
\; \int d^{3} r_{p} \; \phi_{\rm int} ({\bf r}_{e} - {\bf r}_{p})
\phi_{\rm int}^{\ast} ({\bf r}'_{e} - {\bf r}_{p})   / \Omega 
\nonumber \\
&=& \exp \left( i \frac{m_{e}{\bf P}}{\hbar M} \cdot
[{\bf r}_{e} - {\bf r}'_{e} ] \right)  \rho_{\rm int} ({\bf r}_{e},
{\bf r}'_{e})
\label{e:rhoe}
\end{eqnarray}
where we introduce the electron density matrix $\rho_{\rm int}$
for electrons in atoms at rest $({\bf P} = 0)$.
	It is interesting to note that the electron subsystem
of the hydrogen atom
is homogeneous.  To prove that, it is sufficient to show
that $\rho_{\rm int}$ depends on ${\bf r}_{e}$ and ${\bf r}'_{e}$
as ${\bf r}_{e} - {\bf r}'_{e}$. Substituting 
${\bf r}'_{e} - {\bf r}_{p}
= {\bf y}$ and writing the argument of $\phi_{\rm int}$,
${\bf r}_{e} - {\bf r}_{p}$, as ${\bf r}_{e}-{\bf r}'_{e}+{\bf y}$, 
we do indeed find that
\begin{equation}
\int d^{3} r_{p} \; \phi_{\rm int} ({\bf r}-{\bf r}_{p})
\phi^{\ast}_{\rm int}({\bf r}'_{e} - {\bf r}_{p})
= \int d^{3} y \; \phi_{\rm int} ({\bf r}_{e} - {\bf r}'_{e} + {\bf y}) 
\phi^{\ast}_{\rm int} ({\bf y}) \; \; ,
\label{e:rel}
\end{equation}
which depends solely on ${\bf r}_{e} - {\bf r}'_{e}$.
Consequently, the reduced electron 
density matrix (\ref{e:rhoe}) only depends on the relative
position, indicating a homogeneous subsystem.  This might
appear surprising: we do not think usually of the electron
in hydrogen as a homogeneous system.  However, the reduced density matrix 
describes the observation of
hydrogenic electrons in the assumption that the proton is `invisible'.
Furthermore Eq.({\ref{e:psrerp}) describes a hydrogen of fixed
momentum (e.g. a beam of hydrogen atoms of well-defined velocity) 
so that the center-of-mass position of the atom is undetermined.
It is then equally likely to observe the electron in any position,
and the electron subsystem is homogeneous.

	Consequently, as discussed in the previous section,
the Fourier-transform of the density-matrix plays a central role. 
The density-matrix $\rho_{\rm int}$ of a $1s$-electron 
in a hydrogen atom at rest
(${\bf P} = 0$ in Eq.(\ref{e:rhoe})),
has a simple analytical Fourier-transform,
\begin{equation}
\tilde{\rho}_{\rm int} ({\bf k}) =
\frac{1}{\Omega} \; \frac{64 \pi a_{0}^{3}}{1 + (a_{0} k)^{2} } \; \; .
\label{e:rhoint}
\end{equation}
The Fourier transform $\tilde{\rho}$
for an electron in an atom of arbitrary momentum ${\bf P}$
is a translation in ${\bf k}$-space
of $\tilde{\rho}_{int}$.  To show that, we insert the inverse Fourier
transform  of $\tilde{\rho}$ in the expression for the
density matrix (\ref{e:rhoe}),
\begin{equation}
\rho({\bf r}_{e},{\bf r}'_{e}) =
\frac{1}{(2\pi)^{3}} \; \int d^{3} k' \; \tilde{\rho}_{\rm int}({\bf k}')
\exp \left( i \left[ \frac{m_{e} {\bf p}}{M \hbar} + {\bf k}' \right]
\cdot [{\bf r}_{e} - {\bf r}'_{e}]
\right) \; \; .
\label{e:rhoef}
\end{equation}
The substitution, ${\bf k} = {\bf k}' + \frac{ m_{e} {\bf P}}
{M \hbar}$, then leads to the Fourier-transform:
\begin{equation}
\rho({\bf r}_{e},{\bf r}'_{e}) = 
\frac{1}{(2\pi)^{3}} \; \int d^{3} k \; \tilde{\rho}_{\rm int} \left( {\bf k} - 
\frac{m_{e} {\bf P}}{M\hbar} \right) \exp(i{\bf k} 
\cdot [{\bf r}_{e} - {\bf r}'_{e}]) \; \; ,
\label{e:rhoin}
\end{equation}
from which it follows that 
\begin{equation}
\tilde{\rho}({\bf k})
= \tilde{\rho}_{\rm int} ({\bf k} - \frac{m_{e} {\bf P}}{M\hbar}) \; \; .
\label{e:rhot}
\end{equation}
Thus, compared to $\tilde{\rho}_{\rm int}$,
$\tilde{\rho}$ is displaced in ${\bf k}$-space by
$\frac{m_{e} {\bf P}}{M\hbar}$.  
Similarly, the momentum distribution 
$f({\bf p}) = [\Omega/(2\pi \hbar)^{3}] \tilde{\rho}_{int}({\bf p}/\hbar)$,
is related to the electron momentum distribution $f_{\rm int}({\bf p})$
for atoms at rest, by means of a simple displacement in momentum space:
\begin{equation}
f({\bf p}) = f_{\rm int} ({\bf p} - \frac{m_{e}}{M} {\bf P}) \; \; \; ,
\label{e:ftr}
\end{equation}
where 
\begin{equation}
f_{\rm int} ({\bf p}) = \frac{1}{(2\pi \hbar)^{3}} 
\frac{64 \pi a_{0}^{3}}{1 + \left( \frac{a_{0} p}{\hbar} \right) } \; \; \; .
\label{e:ftrint}
\end{equation}
Equation (\ref{e:ftr}) expresses a Gallilean transformation
from the center-of-mass reference frame (${\bf P} = 0$) to the lab frame
of reference in which the atom is moving with velocity
${\bf P}/M$.  If the electron in the center-of-mass frame has velocity
${\bf v} = {\bf p}/m_{e}$, then, in the lab frame, its velocity
is observed as ${\bf v}' = {\bf p}/m_{e} - {\bf P}/M$, corresponding
to a momentum equal to $m_{e} {\bf v}' = {\bf p} - \frac{m_{e}}{M} {\bf P}$.

	Since $f$ and $f_{\rm int}$ are equal up to a translation in
momentum space, their widths, or standard deviations, are equal.
In the center-of-mass frame,
$\langle {\bf p} \rangle = 0$, so that the standard deviation $\Delta p
= \sqrt{ \langle {\bf p}^{2} \rangle - (\langle {\bf p} \rangle)^{2} }$
is equal to
\begin{eqnarray}
\Delta p &=& \sqrt{ \int d^{3} p \; p^{2} f_{\rm int}({\bf p}) } 
\nonumber \\
&=& \frac{\hbar}{a_{0}} \; \; \; \; \; .
\label{e:fin}
\end{eqnarray}
Thus, although the electron subsystem is homogeneous, as a consequence
of the electron-proton correlation, a measurement
of the electron momentum can yield a finite range of values.  
In accordance with the Heisenberg uncertainty principle (the correlations
confine the electron in the center-of-mass frame to a region of size
$\sim a_{0}$) the size of this region in momentum space,
or more precisely, the standard
deviation $\Delta p$ of the momentum distribution, is equal $\frac{\hbar}{a_{0}}$.
In the context of describing correlating systems, 
$\Delta p$ is a measure of the strength of the correlation: the
stronger the electron and proton are correlated, the higher the value
of $a_{0}$ and the larger the region in momentum space ($\Delta p$) over
which the electron momentum is `spread out'.

\section{Conclusion}

In this paper, we have reviewed aspects of density matrix theory which
pertain to the description of the entanglement of correlated subsystems.  
The mutual
entanglement of subsystems of a larger (pure) system is conveniently
quantified by the standard deviation of the reduced density matrix
eigenvalues of the entangled subsystems.  For the simple (but relevant) 
case of homogeneous single particle subsystems, 
the density matrix eigenvalues give the particle's
momentum distribution.  Thus, the standard deviation of the momentum
distribution is a measure of the entanglement of the single particle
system, as
we have illustrated for the specific example of the electron in
the two-body hydrogen atom.

\section*{Acknowledgments}
One of the authors, E. T.,  acknowledges support by the NSF 
through a grant for the
Institute for Atomic and Molecular Physics at Harvard University
and Smithsonian Astrophysical Observatory.

\end{document}